# Theory of low power ultra-broadband terahertz sideband generation in bi-layer graphene

J. A. Crosse[1], Xiaodong Xu[2], Mark S. Sherwin[3], & R. B. Liu[1]*

[1] *Department of Physics and Center for Quantum Coherence, The Chinese University of Hong Kong, Shatin, New Territories, Hong Kong, China.*

[2] *Department of Physics, Department of Material Science and Engineering, University of Washington, Seattle, Washington 98195, USA.*

[3] *Department of Physics and the Institute for Terahertz Science and Technology, University of California at Santa Barbara, Santa Barbara, California 93106, USA.*

\* e-mail: rbliu@phy.cuhk.edu.hk

**Abstract**

In a semiconductor illuminated by a strong terahertz field, optically excited electron-hole pairs can recombine to emit light in a broad frequency comb evenly spaced by twice the terahertz frequency. Such high-order terahertz sideband generation is of interest both as an example of extreme nonlinear optics and also as a method for ultrafast electro-optical modulation. So far, this phenomenon has only been observed with large field strengths (~10 kVcm$^{-1}$), an obstacle for technological applications. Here we predict that bi-layer graphene generates high-order sidebands at much weaker terahertz fields. We find that a terahertz field of strength 1 kVcm$^{-1}$ can produce a high-sideband spectrum of about 30 THz, 100 times broader than in GaAs. The sidebands are generated despite the absence of classical collisions, with the quantum coherence of the electron-hole pairs enabling recombination. These remarkable features lower the barrier to desktop electro-optical modulation at terahertz frequencies, facilitating ultrafast optical communications.

**Introduction**

The optical modulation that occurs when an intense light field interacts with matter has a wide range of applications, from molecular imaging [1, 2] to soft X-ray generation [3-5] and potential applications in all-optical signal processing. The well-known process of high-harmonic generation (HHG) [6] - where the interaction of an optical field with an atom results in the emission light in a broad frequency comb evenly spaced by twice the optical field frequency - is an example of such a process.

HHG can be understood with a three step model. In the first step an electron is ionized from an atom by an intense field. In the second, the ionised ionized electron is accelerated by the field along a closed orbit, returning to the ion at a later time with a higher energy. Finally, the electron recombines with the ion releasing a photon which is an odd harmonic of the original field. One important feature of the process is its `non-perturbative' nature, which results in a `plateau-like' spectrum where the intensity of the harmonics remain approximately constant up to a cutoff frequency of $I_\mathrm{p} + 3.17 U_\mathrm{p}$ (where $I_\mathrm{p}$ is the atom binding energy and $U_\mathrm{p} = \frac{e^2 E^2}{4 m_\mathrm{e} \omega^2}$ is the ponderomotive energy with $E$ and $\omega$ the strength and angular frequency of the light field and $e$ and $m_\mathrm{e}$ the charge and mass of the driven electron, respectively), outside of which the harmonic strength decays away exponentially [7-9].

The similar process of high-order sideband generation (HSG) can occur in semiconductors [10]. In HSG, a weak optical field promotes a valence band electron to the conduction band, creating an electron-hole pair. The pair is then driven in the band by an intense, THz frequency field. In analogy with HHG, the recombination of the electron-hole pair also produces light in a broad frequency comb [Fig. 1(a)]. However, in this case, the resulting sidebands in the excitation field spectrum occur at even harmonics of the THz field. As with HHG, a `non-perturbative' plateau occurs with a cutoff of $\Delta + 3.17 U_\mathrm{p}$ (where $\Delta$ is the detuning of laser from the band gap of the semiconductor and $U_\mathrm{p} = \frac{e^2 E_\mathrm{THz}^2}{4 m_\mathrm{R} \omega_\mathrm{THz}^2}$ is the ponderomotive energy with $E_\mathrm{THz}$ and $\omega_\mathrm{THz}$ the strength and angular frequency of the THz field and $m_\mathrm{R}$ the electron-hole reduced mass). Such high-order THz sideband generation (HSG) is not only of fundamental interest, as an example of extreme nonlinear optics in condensed matter systems, but also of technological importance as a mechanism for ultrafast electro-optical modulation - a crucial ingredient in Tbit s$^{-1}$ rate optical communication. Hence, HSG has a wide range of potential applications but there is a serious technological challenge; HSG requires large THz field strengths. Current experimental observations of HSG have employed THz field amplitudes on the order of 10 kVcm$^{-1}$ [11] generated by a free-electron laser [12, 13]. The need for such large field strengths presents an obstacle to the use of HSG in optical devices.

Novel materials, such as graphene, offer a potential solution to this problem. The maximum modulation bandwidth is given by the maximum kinetic energy that can be obtained from the THz field by the electron-hole pair. This, in turn, is given by the work done on the electron-hole pair by the THz field, $\int e E_\mathrm{THz} v \, dt$, which is linearly proportional to their velocity, $v$. Electrons in graphene act like massless Dirac electrons and hence have a large initial velocity ($v_\mathrm{f} \approx 10^6$ ms$^{-1}$). Conversely, electrons in conventional parabolic band semiconductors have vanishing initial velocity near the band edge (although such velocities are available away from the band edge). Hence, graphene's linear dispersion and ballistic electrons offer a significant energy gain, at least at low energies, over the parabolic bands of traditional semiconductor and, therefore, is a promising candidate material for HSG at the low-field intensities which would be suitable for optical device design. In addition, graphene's resonant transition at telecomm frequencies [14], weak uniform absorptive loss [15] and ultrafast optical response [16] make it an ideal candidate for optical communication applications. In fact, it has been demonstrated that graphene, owing to its ultrafast electrical response, is a promising material for ultrafast electro-optical modulation [17].

Here we report on the prediction that, although conversion efficiency is much lower, bi-layer graphene (BLG) [18-20] produces a much broader bandwidth of modulation than conventional semiconductors and, specifically, BLG is a versatile system for HSG at telecommunication wavelength (1.55 μm). We find that a THz field of strength 1 kVcm$^{-1}$ can produce an HSG spectrum of about 30 THz in the telecom waveband, 100 times broader than in GaAs (or, equivalently, a comparable bandwidth in GaAs would require THz field intensities that are 100 times stronger). These remarkable features make it possible to realise HSG with a much weaker THz laser, which lowers the barrier to desktop electro-optical modulation at THz frequencies, facilitating ultrafast optical communication.

**Results**

**HSG in the absence of classical collisions**

The analysis of HSG in a conventional semiconductor is semi-classical; the sideband spectrum is found from the optical emission associated with the classical collision of electron-hole pairs that have been driven along closed trajectories [Fig. 1(a)]. From a semi-classical perspective, however, HSG in graphene seems impossible. Optically excited electron-hole pairs, owing to their massless nature, act like photons in that their velocity is fixed at the Fermi velocity. The action of the THz field changes the momentum and energy of the electron and hole but the velocity remains constant. As a result, the electron and hole separate at a constant rate and do not return to the same location [Fig. 1(b)] and, hence, no classical collisions occur. A similar issue is shared by electron-hole pairs excited well above the band edge. In this regime, the carriers, as with graphene, are created with a large initial velocity but, again, the THz field is usually not strong enough to reverse their direction of motion and classical collisions are absent. Recent work has shown that HSG can still be created in the absence of classical collisions if the coherence time of the electron-hole pairs is longer than half the THz period [21] (also see Supplementary Note 1 and Supplementary Fig. 1). The long coherence time leads to broad electron and hole wavefunctions that can overlap significantly, even when the classical separation is non-zero. Hence, the electron and hole can still recombine to form sidebands even in the absence of classical collisions. As the semi-classical collisions are absent in graphene, it is this 'quantum' process that is the dominant mechanism of HSG in this material. Furthermore, the absence of the recollision condition means that the carriers can be accelerated by the THz field for the full half cycle and hence attain an energy that is greater than the $\Delta + 3.17 U_\text{p}$ limit imposed by the necessity of classical recollision. Thus, this quantum coherence induced sideband plateau is broader than the semi-classical plateau.

**HSG in bi-layer graphene**

In the following study the optical spectrum of a THz driven electron-hole pair in biased BLG is investigated. We consider BLG, as opposed to monolayer graphene, since a bias field can be applied to the dual layer structure to open a gap at the Dirac point [22][Fig. 2(a)], which avoids excitations of many electron-hole pairs by the THz field (Schwinger instability [23]), whilst still maintaining linear dispersion at the telecomm frequencies. The low energy band structure of BLG [19, 20] (also see Supplementary Note 2) displays four, valley degenerate bands of which the two of lower energy, $E_\text{v}$ and $E_\text{v'}$, are fully occupied valence bands and the two of higher energy, $E_\text{c}$ and $E_\text{c'}$, are fully empty conduction bands [Fig. 2(b)]. With a relatively low intensity THz field and an excitation field in the C-band of the telecomm frequency range only one of the four possible transitions is significant. The major contribution to the spectrum comes from the $E_\text{v} \rightarrow E_\text{c}$ transition [labelled (4) in Fig. 2(b)] where both the electron and hole are created in the linear-dispersion region. This is where the initial velocity is highest and hence where the maximum kinetic energy gain can be obtained. The other transitions [labeled (1-3) in Fig. 2(b)] contribute to a lesser extent to the HSG spectrum as either the electron or hole is created in the parabolic-dispersion region or the optical excitation is off resonant or both (the HSG spectrum due to transitions 1 and 3 is narrower by approximately a factor of two and the HSG due to transition 2 negligible). Thus, the maximum modulation is determined by $E_\text{v} \to E_\text{c}$ transition alone.

We treat the THz-field as a continuous wave of the form, $\mathbf{F}_{\text{THz}}(t) = \mathbf{F}_{\text{THz}} \cos(\omega_{\text{THz}} t)$, which we take to be polarized in the x-direction. Similarly, the weak excitation field is taken to be a continuous wave, $\mathbf{F}_{\text{ex}}(t) = \mathbf{F}_{\text{ex}} e^{-i\omega_{\text{ex}} t}$. The coupling of the THz field to the material is introduced via the time-dependent canonical momentum by the inclusion of the vector potential $\mathbf{k} \to \widetilde{\mathbf{k}}(t) = \mathbf{k} + \mathbf{A}_{\text{THz}}(t)$ with $\mathbf{F}_{\text{THz}} = -\frac{\partial \mathbf{A}_{\text{THz}}(t)}{\partial t}$. If the excitation field is sufficiently weak one can work in the quasi-equilibrium regime where the depletion of the valence band is negligible. We also note that the large gap induced by the bias field prevents interband transitions caused by the THz field. These facts allow us to work in the single active electron approximation and, hence, ignore electron-electron correlation effects.

**Simulation parameters and results**

We evaluate this system by direct simulation of the time dependent Schrödinger equation for an excitation field of 1.55 μm (194 THz), which is located in the C-band of the telecomm frequencies, and a THz field of frequency 1 THz. The potential induced by the bias voltage on the BLG was set to 0.28 eV, which is achievable with a field of ≈3 V nm$^{-1}$ [22]. The dephasing factor, Γ, is introduced to account for various relaxation mechanisms. Studies of hot carrier relaxation in monolayer graphene and BLG have shown that carrier lifetimes are on the order of 100s of femtoseconds [24-26]. However, these studies have been performed at high pump intensities (e.g. $\approx 6\times 10^5$ mW μm$^{-2}$ [24]). At such high intensities one expects dense carrier concentrations on the order of $10^{13}$ cm$^{-2}$. In this regime fast carrier-carrier scattering will dominate. HSG experiments are conducted at much lower pump intensities ($\approx 3 \times 10^{-4}$ mW μm$^{-2}$ [11]) and the resulting low carrier concentrations will lead to a suppression of the carrier-carrier scattering rate. Simulations that include both carrier-carrier and carrier phonon scattering indicate that at such NIR field intensities the relaxation rate will be on the order of ≈600 fs [See Supplementary Figs. 2 and 3. A detailed discussion of both the numerical simulations for the relaxation rates and the effect of relaxation rates on the HSG spectrum can be found in Supplementary Note 3]. Hence, we set to the relaxation time, $\hbar/\Gamma$, to 658 fs ($\Gamma$=1 meV).

The sideband spectrum was calculated for a number of THz field strengths, ranging from 1 kV cm$^{-1}$ to 4 kV cm$^{-1}$. Figure 3(a) shows the sideband spectrum for increasing THz field strength. At a THz field strength of 1 kV cm$^{-1}$ one can see a plateau, symmetric about the excitation frequency, of width ≈28 THz [Fig 3(b)]. As one increases the THz field strength the plateau grows linearly reaching a width of about 108 THz for a 4 kV cm$^{-1}$ field [Fig 3(c)]. The amplitude of the sidebands is much smaller than that of the excitation field. In conventional semiconductors the sideband amplitude has been measured to be about two orders of magnitude smaller than the excitation field [11]. By comparing simulations of HSG in GaAs quantum wells with the experimental data in Ref. [11], we expect that the high sideband amplitudes in BLG to be as much as six orders of magnitude smaller than the excitation field. This comparative reduction in the sideband amplitude compared to conventional semiconductors near the band edge is unsurprising since the process that generates the sidebands in BLG is a quantum process rather than a semi-classical process.

**Plateau cutoff law for linear band materials**

To further understand the plateau of HSG in BLG, we compare the theoretical model to the equivalent theory for conventional semiconductors. For a parabolic band material the Schrödinger equation can be solved in a semi-classical approximation to give an expression for the plateau cutoff $\Delta + 3.17 U_\text{p}$ [9, 10]. One can follow a similar procedure to obtain a cutoff law for BLG (see Supplementary Note 4). The $E_\text{v}$ and $E_\text{c}$ bands can be expanded to first order about the excitation energy

$$E_\text{c}[\widetilde{\boldsymbol{k}}(t)] - E_\text{v}[\widetilde{\boldsymbol{k}}(t)] \approx E_\text{g} + 2\hbar \widetilde{\mathbf{v}}_\text{f,vc} \cdot (\widetilde{\boldsymbol{k}}(t) - \boldsymbol{k}_\text{ex}). \tag{1}$$

The result is a band structure that resembles a shifted version of monolayer graphene with effective Fermi velocity, $\widetilde{\mathbf{v}}_\text{f,vc}$. Solving a restricted version of the classical equations of motion for such a system [21], one obtains a prediction for the plateau cutoffs of $4\widetilde{U}_\text{p,vc}$ above and below the excitation frequency, with linear ponderomotive energy given by $\widetilde{U}_\text{p,vc} = \frac{e\widetilde{\mathbf{v}}_\text{f,vc} \cdot \mathbf{F}_\text{THz}}{\omega_\text{THz}}$. The dashed black line in Fig. 3(a) shows the theoretical prediction of the cutoff. One observes good agreement over the full range of THz field strengths simulated. Similarly, the black arrows in Figs. 3(b) and (c) mark the theoretical cutoff given by the $4\widetilde{U}_\text{p,vc}$ cutoff law. Figure 3(b) shows that at low field strengths the cutoff law is in good agreement with the numerical results. Figure 3(c) shows that although the cutoff law correctly predicts the high end cutoff it over-estimates the low end cutoff. The reason for this is that as the bands move towards the Dirac point they deviate from the linear model of Eq. (1) and, hence, the electron and hole have lower velocities. This leads to the discrepancy observed in Fig. 3(c).

**Discussion**

The advantages of BLG as a material for HSG can be seen by comparison with conventional semiconductors such as GaAs. In order to obtain a bandwidth of about 28 THz with a 1 THz frequency field one would need a field strength of nearly 14 kV cm$^{-1}$ (i.e., the required THz field intensity for GaAs is ~200 times stronger than for the BLG). Current experiments have used free electron lasers to produce such fields. In principle, such fields can be created in solid state devices by careful microwave engineering [11]. However, the field reduction offered by BLG will make the engineering of such a device easier as well as resulting in a more compact, lower power design. Another advantage is that BLG has resonant transitions throughout the telecomm frequencies (including all standard band classifications from O-band to U-band, not just the conventional C-band "erbium window"), thus allowing broadband modulation at any frequency in this range. This would mean that BLG based devices could operate optimally over a wide range of frequencies without the need to further tailor the material. Conversely, the GaAs bandgap (which corresponds to a wavelength of ~0.87 μm) is outside the telecomm frequency waveband. Even materials with smaller band gaps (e.g. silicon ~1.12 μm, germanium ~1.85 μm) usually need further engineering (either by alloying or via the creation of heterostructures) in order to create a device that is optimally tuned to a particular optical frequency. For exceptionally small band-gap semiconductors (such as InAs ~3.44 μm), a field in the telecomm waveband can also create electron-hole pairs well above the band edge and therefore with high initial velocity. However, the carrier relaxation in such materials, due to the rapid emission of LO phonons (which have energies of tens of meV), would be much faster than in graphene (where the LO phonon energy ~200 meV [27]).

For optical device applications one needs to consider the depth of the modulation. Again, the versatile properties of BLG are a bonus and a number of techniques can be employed to increase the amplitude of the sidebands. For example, employing waveguides to increase the interaction length between BLG and the laser, would increase the sideband amplitude and hence the modulation depth. Furthermore, owing to the small reflectance of BLG (<0.1%) [15] and the fact that the modulated light is emitted equally in the forward and backward directions, an enhanced modulation depth can be obtain by detection in reflection rather than in transmission. In addition, the compact nature of the process and the fast response time of the effect would both be great advantages of a BLG modulator.

The results of this study suggest that biased BLG can provide ultra-broad HSG and, hence, ultrafast modulation of an optical field at THz field strengths low enough for table top sources to be used [28-30]. These results open up the way for ultrafast, HSG based, optical devices, which, coupled with the remarkable properties of BLG, marks a step towards realising all optical broadband modulation and optical communication at the Tbit s[-1] rate.

**Methods**

**Time-dependent Schrödinger equation**

The electron-hole pair wavefunction, $P_{\sigma,\lambda}(\widetilde{\boldsymbol{k}}, t) = \langle \hat{e}_{\sigma,\lambda}(\widetilde{\boldsymbol{k}}, t)\hat{h}_{\sigma,\lambda}(\widetilde{\boldsymbol{k}}, t)\rangle$, where λ and σ are the spin and valley indices respectively, was computed using the time-dependent Schrödinger equation

$$i\hbar\frac{\partial}{\partial t}P_{\sigma,\lambda}(\widetilde{\boldsymbol{k}}, t) = \Delta E[\widetilde{\boldsymbol{k}}(t)]P_{\sigma,\lambda}(\widetilde{\boldsymbol{k}}, t) + i\mathbf{d}_{vc}[\widetilde{\boldsymbol{k}}(t)] \cdot \mathbf{F}_{ex}(t), \quad (2)$$

with

$$\Delta E[\widetilde{\boldsymbol{k}}(t)] = E_c[\widetilde{\boldsymbol{k}}(t)] - E_v[\widetilde{\boldsymbol{k}}(t)] - \hbar\omega_{ex} - i\Gamma + i\{\mathbf{A}_c^*[\widetilde{\boldsymbol{k}}(t)] + \mathbf{A}_v[\widetilde{\boldsymbol{k}}(t)]\} \cdot \mathbf{F}_{THz}(t).$$

(3)

Here, $E_c[\widetilde{\boldsymbol{k}}(t)]$ and $E_v[\widetilde{\boldsymbol{k}}(t)]$ are the instantaneous valence and conduction band energies, which can be derived using a tight-binding method with nearest neighbour hopping [18-20].

The time-dependent canonical momentum is given by $\widetilde{\mathbf{k}}(t) = \left(\mathbf{k} - \frac{e}{\hbar}\int dt'\, \mathbf{F}_{\text{THz}}(t')\right)$, and $\mathbf{d}_{\text{vc}}[\widetilde{\mathbf{k}}(t)] = e\langle v, \widetilde{\mathbf{k}}(t)|\nabla_{\mathbf{k}}|c, \widetilde{\mathbf{k}}(t)\rangle$ and $\mathbf{A}_i[\widetilde{\mathbf{k}}(t)] = e\langle i, \widetilde{\mathbf{k}}(t)|\nabla_{\mathbf{k}}|i, \widetilde{\mathbf{k}}(t)\rangle$ are the interband dipole moment and Berry connection terms [31], the latter of which is required to preserve gauge invariance. As the driving of the electrons is linear (as opposed to closed orbits) the contribution of the Berry connection term is negligible (see Supplementary Note 5). $\Gamma$ is a phenomenological dephasing rate that accounts for various relaxation mechanisms. For the sake of simplicity, we have not included the Coulomb interaction. This is justified by the condition that the resonant excitation is taken to be far from the band edge and hence off-resonant with the bound state excitations. Furthermore, at such high quasi-momenta, the kinetic term in the Hamiltonian will dominate. Equation (2) was evaluated directly using standard numerical techniques. The resulting polarization is given by

$$\mathbf{P}(t) = i\langle \widehat{\mathbf{d}}_{\text{vc}}(t)\rangle = i\sum_{\lambda,\sigma} \int d^3k\, \mathbf{d}^*_{\text{vc}}[\widetilde{\mathbf{k}}(t)]P_{\sigma,\lambda}(\widetilde{\mathbf{k}},t), \qquad (4)$$

with the sideband spectrum calculated via discrete Fourier transform.

**Full methods and supplementary figures are available online.**

**References**

1. Itatani, J., et al. Tomographic imaging of molecular orbitals. *Nature* **432**, 867-871 (2004).
2. Baker, S. et al. Probing proton dynamics in molecules on an attosecond time scale. *Science* **312**, 424-427 (2006).
3. Spielmann, Ch., et al. Generation of coherent x-rays in the water window using 5-femtosecond laser pulses. *Science* **278**, 661-664 (1997).

**Acknowledgements** We acknowledge supports from Hong Kong Research Grants Council – General Research Fund Project 401512 and The Chinese University of Hong Kong Focused Investments Scheme.

**Author contributions** R.B.L. conceived the idea and supervised the project. R.B.L., X.D.X., and M.S. designed the device. J.A.C. carried out the theoretical study and numerical calculation. J.A.C. wrote the paper. All authors analysed the data and commented on the manuscript.

**Additional information** The authors declare no competing financial interests. Supplementary information accompanies this paper. Correspondence and requests for materials should be addressed to R.B.L.

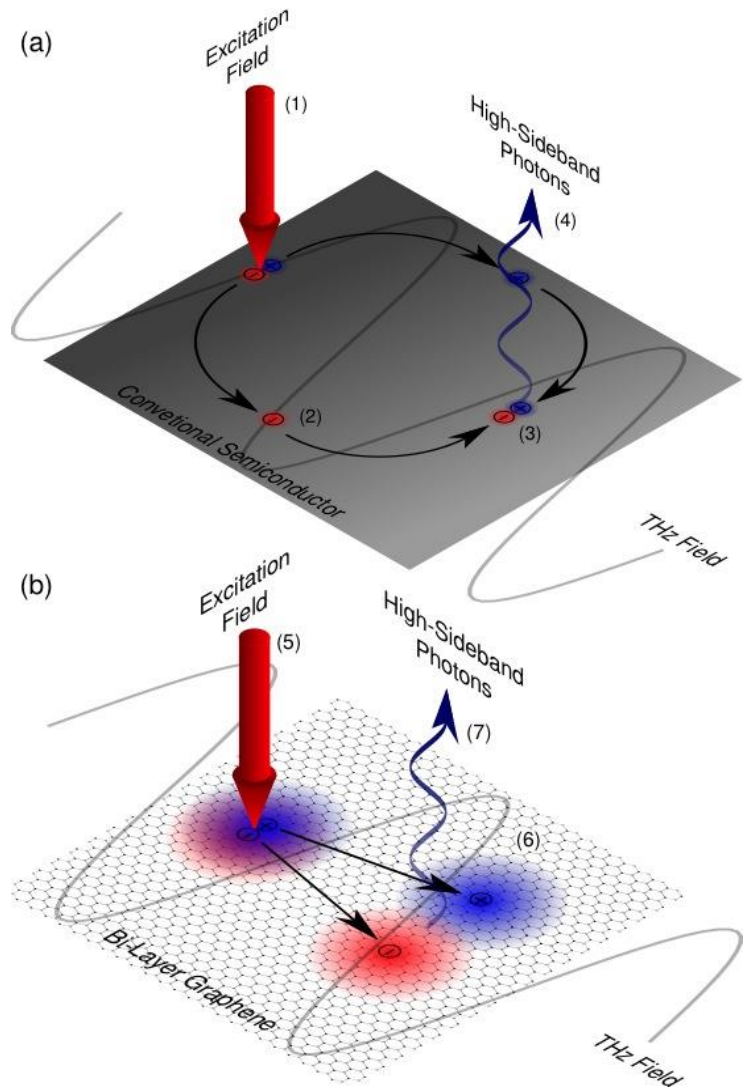

**Figure 1: Schematics of high-sideband generation in conventional semiconductors and in graphene. (a)** In a conventional semiconductor, an electron-hole pair, created by a weak excitation field (1), is accelerated by an intense THz frequency field gaining energy. The electron and hole wavefunctions separate completely (2). The THz field drives the electron and hole back towards each other, recombining (3) and releasing a high-sideband photon (4). **(b)** In graphene, a weak excitation field excites an electron-hole pair (5) which is accelerated in opposite directions by the THz field (6). As the electron-hole pair has vanishing effective mass the THz alters their energy and momentum but not their velocity. Hence the electron and hole separate at the Fermi velocity. However, owing to the broad wavefunction

of the electron and hole, there is still significant overlap at finite separations and, hence, recombination (and the release of a high-sideband photon) is possible even though the classical locations of the electron and hole do not coincide (7).

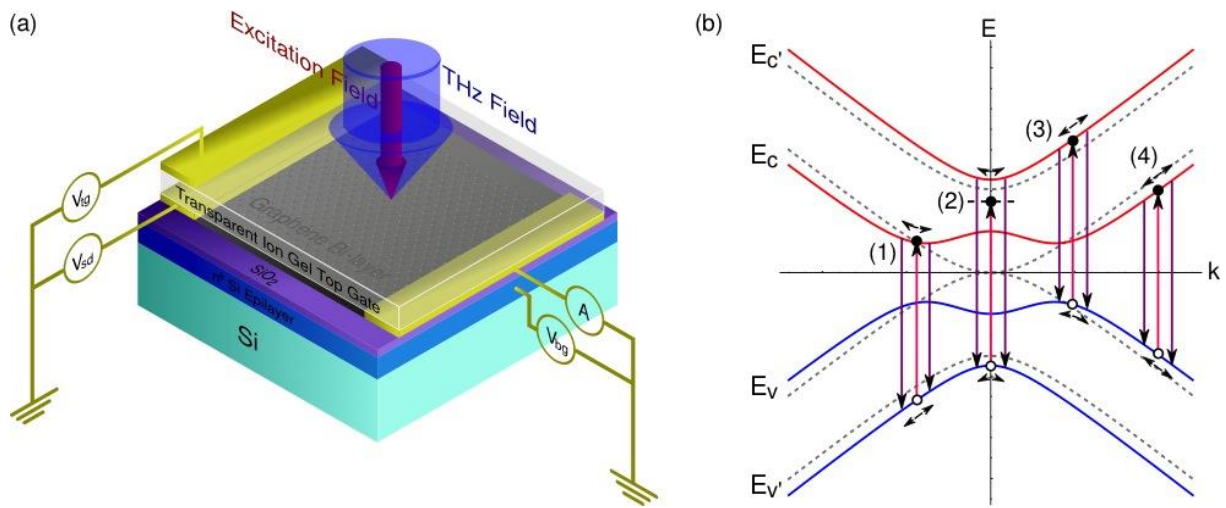

**Figure 2: Device structure and mechanism for high-sideband generation in bi-layer graphene. (a)** Device design for high-sideband generation. A single sheet of BLG is layered on a silicon dioxide substrate and biased via a gate voltage. The top gate is transparent in order to allow the incident THz and optical excitation fields to reach the graphene layer. The n⁺ Si epilayer should be thin compared to the penetration depth of the THz field in order to avoid significant reflection. **(b)** Acceleration of electrons/holes along the energy bands. The grey curves show the band structure for unbiased BLG. The four hyperbolic bands are gapless at the Dirac point rendering the material a semi-metal. The application of a bias field opens a gap at the Dirac point, converting BLG from a semi-metal to a semi-conductor. The 1.55µm excitation field creates electron-hole pairs, via three resonant [(1), (3) and (4)] and one off-resonant transitions (2). These electron hole pairs are then driven in the band, recombining to produce the optical sideband spectrum. The dominant contribution is from the transition between the $E_\mathrm{v} \rightarrow E_\mathrm{c}$ (4) which occurs in the linear region of the bands.

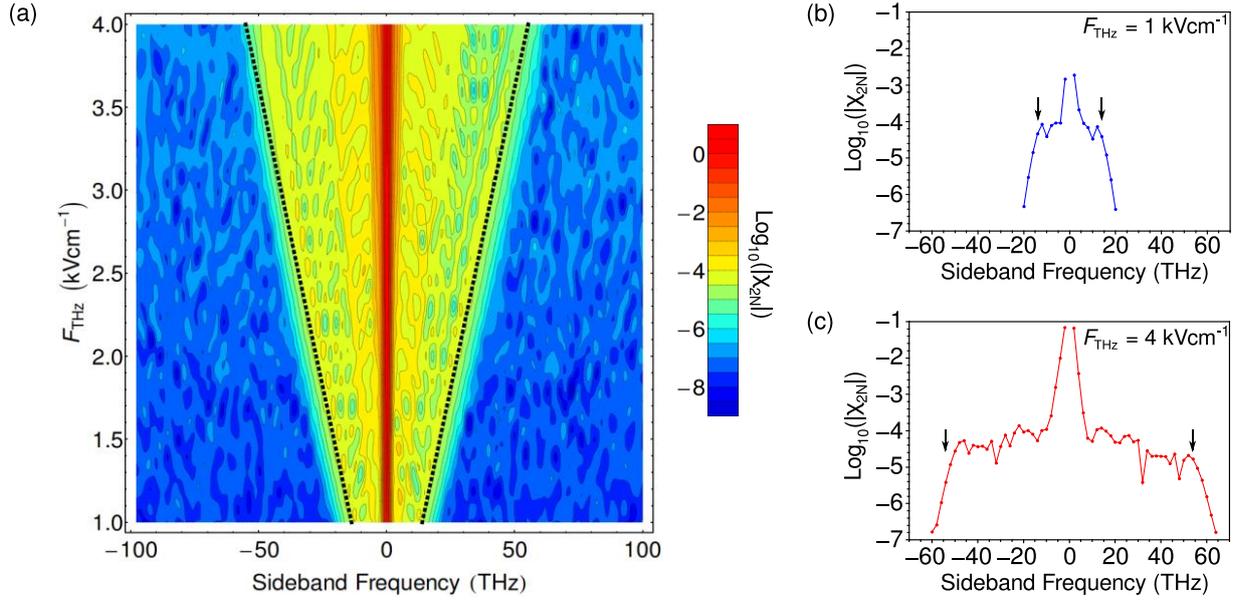

**Figure 3: Simulated data of the sideband spectrum of a THz driven bi-layer graphene. (a)** Contour plot of the sideband spectrum of THz driven BLG for different THz field strengths. The colour map gives the logarithm of the normalized sideband amplitude $\log_{10}(|\chi_{2N}|)$. The dashed black lines give the theoretical sideband cutoffs, $\pm 4\widetilde{U}_{p,vc}$. **(b)** The sideband spectrum for a THz field of frequency 1 THz and strength of 1 kV cm$^{-1}$. **(c)** The sideband spectrum for a THz field of frequency 1 THz and strength 4 kVcm$^{-1}$. The arrows in (b) and (c) mark the location of the theoretical sideband cutoffs, $\pm 4\widetilde{U}_{p,vc}$.

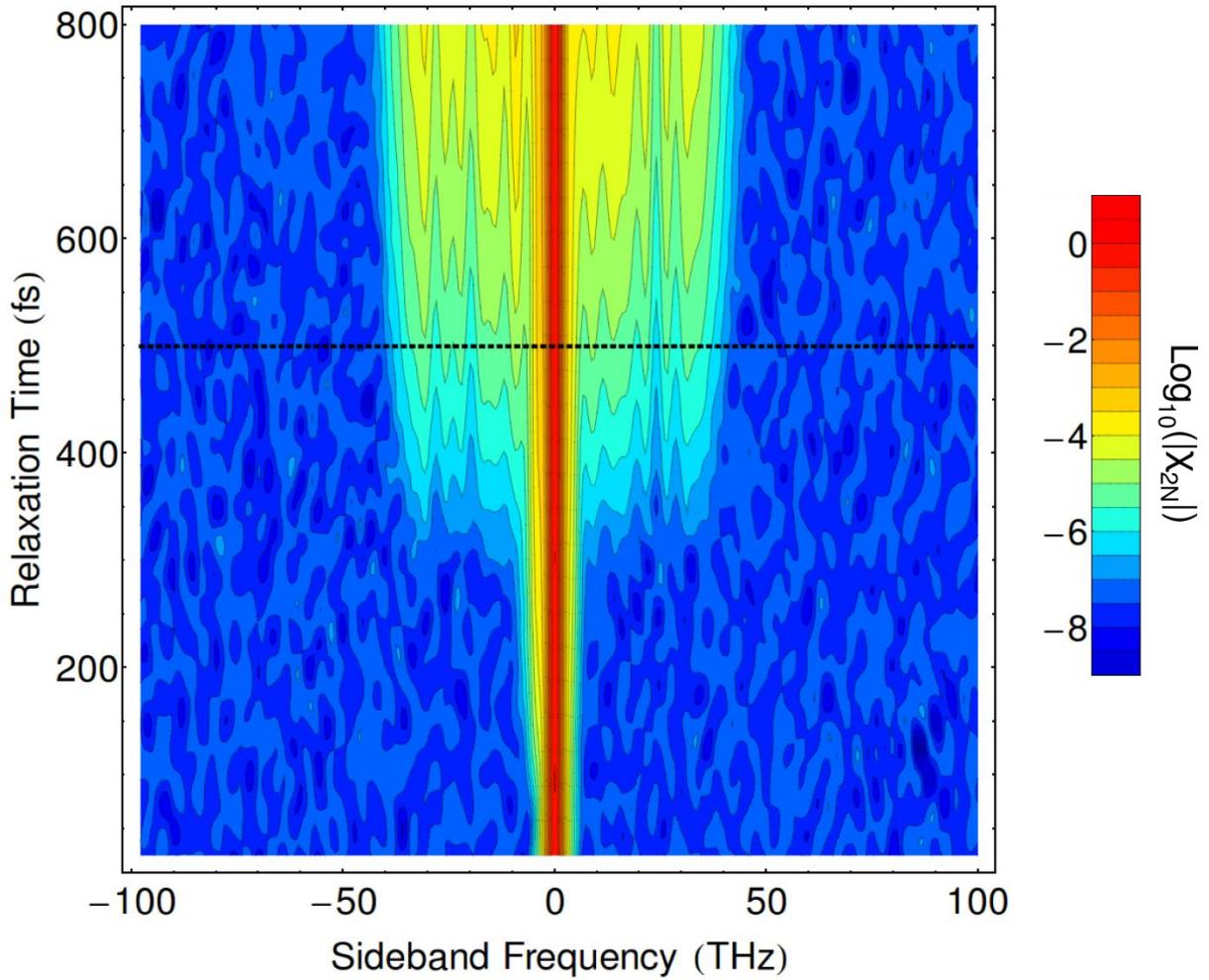

**Supplementary Figure 1: The effect of relaxation times on the sideband spectrum of a THz driven bi-layer graphene.** Contour plot of the sideband spectrum of THz driven BLG for different relaxation times. The colour map gives the logarithm of the normalized sideband amplitude $\log_{10}(|\chi_{2N}|)$. The dashed black line indicates half the THz field period, 500 fs. The strength of the THz field is 2.5 kVcm$^{-1}$.

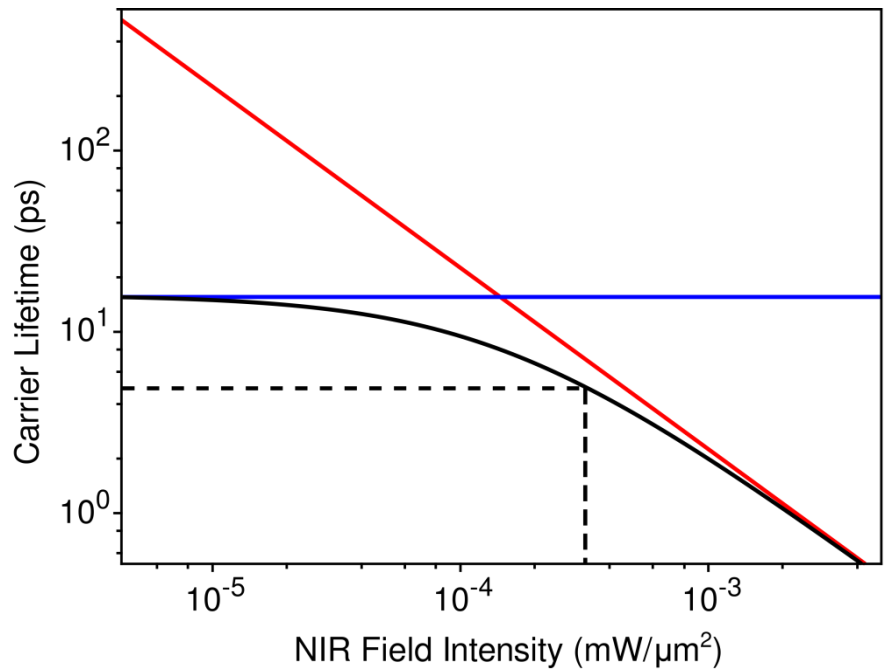

**Supplementary Figure 2: Carrier-carrier scattering time as a function of the NIR field intensity.** The red line is the intraband carrier-carrier scattering rate, the blue line is the impact ionization rate and the black line is the combined scattering rate. The dashed lines mark the typical NIR field intensity used in HSG experiments and the corresponding scattering time.

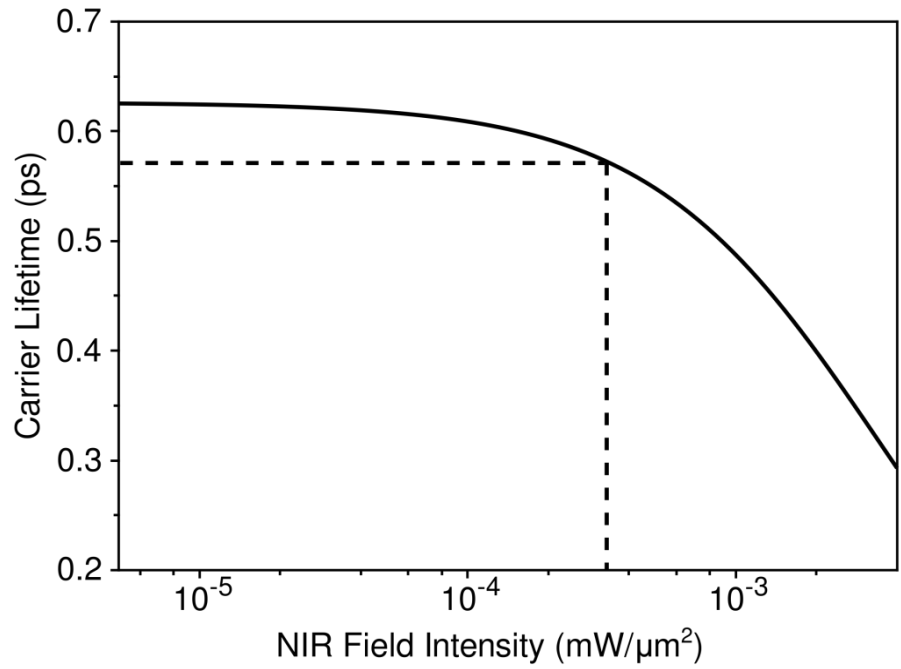

**Supplementary Figure 3: Total (carrier-carrier + carrier-phonon) scattering rate as a function of NIR field intensity.** The dashed lines mark the typical NIR field intensities used in HSG experiments and the corresponding scattering time.

**SUPPLEMENTARY NOTE 1: The effect of relaxation on bandwidth**

The maximum sideband energy for an electron-hole pair that is created at $t - \tau$ and recombines at $t$ is given by

$$\hbar\omega_s = 2\tilde{U}_{p,vc}(\sin[\omega_{THz}(t-\tau)] - \sin[\omega_{THz}t]), \qquad (1)$$

which is maximized for $t = \frac{3\pi}{2}$, $\tau = \pi$ and minimized for $t = \frac{\pi}{2}$, $\tau = \pi$. Thus, in order to obtain the maximum modulation the electron-hole pairs must travel for a full half cycle, which, for the 1 THz driving field described in the main text, is 500 fs. For relaxation times greater than half the period of the THz driving field the electron-hole pair can travel for the necessary length of time for the maximum modulation to occur and, hence, one should observe the full sideband plateau. For relaxation times less than half the time period then electron-hole pairs will relax before they can travel for the necessary length of time to achieve maximum modulation and, hence, the plateau should be absent. One should also observe an increase in the sideband amplitude relative to the NIR excitation field for increasing relaxation time.

Supplementary Fig. 1 shows the effect of the relaxation time on the sideband spectrum of BLG driven by a 2.5 kVcm$^{-1}$ strength field with frequency of 1 THz. The dashed line shows the half cycle time of 500 fs. Above this line, when the relaxation time is longer than the half-cycle time one sees a prominent plateau. Below the line, when the relaxation time is shorter than the half-cycle time, the plateau strength reduces and for very short relaxation times vanishes completely. Note that by increasing the frequency of the THz field one would reduce the half-cycle time and hence reduces the minimum relaxation required to observe a plateau. However, as the linear ponderomotive energy is inversely proportional to the THz field frequency this would lead to a smaller plateau and subsequently a smaller bandwidth modulation.

**SUPPLEMENTARY NOTE 2: The band structure of bi-layer graphene**

The low energy band structure of bi-layer graphene (BLG) can be derived via a tight-binding method with nearest neighbour hopping [1, 2]. This results in two Hamiltonians, one each for the $+$ and $-$ valleys

$$H_{\pm} = \begin{pmatrix} V & v_f k_{\mp} & 0 & 0 \\ v_f k_{\pm} & V & 2\gamma & 0 \\ 0 & 2\gamma & -V & v_f k_{\mp} \\ 0 & 0 & v_f k_{\pm} & -V \end{pmatrix}, \qquad (2)$$

where $k_{\pm} = k_x \pm i k_y$, $v_f = -6.462 \frac{eV\text{Å}}{\hbar}$ is the Fermi velocity [3, 4] (from numerical parameters given in [5]), $\gamma = 0.202\ eV$ is the interlayer hopping [6] (Note that $\gamma$ has been rescaled by a factor of $1/2$ compared with the value in [6]) and $V$ is the bias energy applied to each layer (which, in the main text, is taken to be 0.2 eV, achievable with a bias field of $\approx 3$ Vnm$^{-1}$ [7]). Diagonalization leads to four spin and valley degenerate eigenbands centred about the Fermi energy

$$E_{c'}(\mathbf{k}) = \sqrt{v_f^2 |\mathbf{k}|^2 + 2\gamma^2 + V^2 + 2\sqrt{v_f^2 |\mathbf{k}|^2 (\gamma^2 + V^2) + \gamma^4}}, \qquad (3a)$$

$$E_c(\mathbf{k}) = \sqrt{v_f^2 |\mathbf{k}|^2 + 2\gamma^2 + V^2 - 2\sqrt{v_f^2 |\mathbf{k}|^2 (\gamma^2 + V^2) + \gamma^4}}, \qquad (3b)$$

$$E_v(\mathbf{k}) = -\sqrt{v_f^2 |\mathbf{k}|^2 + 2\gamma^2 + V^2 - 2\sqrt{v_f^2 |\mathbf{k}|^2 (\gamma^2 + V^2) + \gamma^4}}, \qquad (3c)$$

$$E_{v'}(\mathbf{k}) = -\sqrt{v_f^2 |\mathbf{k}|^2 + 2\gamma^2 + V^2 + 2\sqrt{v_f^2 |\mathbf{k}|^2 (\gamma^2 + V^2) + \gamma^4}}. \qquad (3d)$$

In the absence of the bias field, the band structure consists of four hyperbolic bands with a gapless transition at the two Dirac points. As with monolayer graphene (MLG), the Fermi surface passes through the Dirac points resulting in two fully filled valence bands and two fully empty conduction bands. The application of a bias field shifts the bands, converting

BLG from a gapless semi-metal to a gapped semiconductor. For equal and opposite potentials applied to each of the two graphene layers the Fermi level is unchanged in relation to the bands and remains in the gap [See Fig. 2(b) in the main text].

**SUPPLEMENTARY NOTE 3: Relaxation rates in bi-layer graphene**

The relaxation rate of carriers in bi-layer graphene can be modeled using a Fermi's Golden rule approach [8]. The total relaxation rate has two main contributions, namely, carrier-carrier scattering and carrier-phonon scattering. Relaxation of carriers via electron-hole recombination occurs on time scales that are many orders of magnitude greater than the scattering time [9] and hence the contribution of this process to the relaxation rate is negligible.

**Carrier-carrier scattering rate:**

Carrier-carrier scattering occurs owing to the coulomb interaction between the charged carriers. The interaction Hamiltonian for the process reads

$$\widehat{H}_{cc} = \frac{1}{2} \int d^2k\, d^2k'\, d^2q\, V_\mathbf{q} \left\{ \hat{e}^\dagger_{\mathbf{k+q}} \hat{e}^\dagger_{\mathbf{k'-q}} \hat{e}_\mathbf{k} \hat{e}_{\mathbf{k'}} + \hat{h}^\dagger_{\mathbf{k+q}} \hat{h}^\dagger_{\mathbf{k'-q}} \hat{h}_\mathbf{k} \hat{h}_{\mathbf{k'}} - 2\hat{e}^\dagger_{\mathbf{k+q}} \hat{h}^\dagger_{\mathbf{k'-q}} \hat{e}_\mathbf{k} \hat{h}_{\mathbf{k'}} \right\}\bigg|_{\mathbf{q}\neq 0},$$

(4)

where

$$V_\mathbf{q} = \frac{2\pi e^2}{\varepsilon_0 |\mathbf{q}|},$$

(5)

is the 2D coulomb potential in momentum space with matrix elements reading [3, 8, 10]

$$|\langle \widehat{H}_{cc} \rangle|^2 = |V_\mathbf{q}|^2 [1 + \cos(\theta_{\mathbf{k+q}} - \theta_\mathbf{k})][1 + \cos(\theta_{\mathbf{k'-q}} - \theta_{\mathbf{k'}})],$$

(6)

where the trigonometric factors appear owing to the chirality of the carriers. In the following we will assume that the excited carrier density is low and, as such, the occupation of each momentum state is small. Hence Pauli blocking is negligible. Thus, the scattering rate for a given occupied momentum state **k** reads

$$\Gamma_{\mathbf{k}}^{cc} = \frac{4\pi\rho}{\hbar} \int d^2\mathbf{k}' d^2\mathbf{q} \left|\langle \hat{H}_{cc} \rangle\right|^2 |\psi(\mathbf{k}')|^2 \delta[E(\mathbf{k}+\mathbf{q}) + E(\mathbf{k}'-\mathbf{q}) - E(\mathbf{k}) - E(\mathbf{k}')], \qquad (7)$$

with $\rho$ being the 2D carrier density and the $\delta$-function enforcing energy conservation. Here, $|\psi(\mathbf{k}')|^2$ is the probability that momentum state $\mathbf{k}'$ is occupied. This factor can be found from the wavefunction, which is given by integrating solutions to the Schrödinger equation for the electron-hole pair wave function [Eq. (2) in the main text] over one THz field cycle. The extra factor of two comes from the spin degree of freedom (intervalley scattering requires high momentum exchange and hence is suppressed). Supplementary Fig. 2 shows the carrier lifetime as a function of NIR field intensity. The carrier concentration is proportional to NIR field intensity (in the low-concentration regime where the Coulomb screening, Pauli blocking, and impact ionization are negligible) and is calculated by considering the photon flux through the BLG sheet with an external quantum efficiency of 6%, appropriate for the double layer nature of BLG. One sees that the carrier lifetime displays an inverse proportionality to the NIR field intensity as expected. In principle, interband scattering of high energy carriers can induce impact ionization (creation of an electron-hole pair by reducing the energy of an electron or hole) and contribute carrier concentration. This effect, however, is largely suppressed in the biased BLG by the finite energy gap (see detailed discussion below). One should also note that at very high carrier concentrations Pauli blocking and Coulomb screening effects will cause the scattering rate to flatten off, hence the carrier lifetime will be longer than predicted from the inverse proportionality given here.

Current HSG experiments use NIR field intensities of $3.2 \times 10^{-4}$ mWµm$^{-2}$ [11] (indicated by the vertical line in Supplementary Fig. 2), which leads to a carrier lifetime of ≈7.05 ps.

**Impact Ionization:**

The process of impact ionization occurs when a carrier with an energy exceeding the energy of the gap scatters off a valence band electron in such a way that the valence band electron is excited to the conduction band. This process not only contributes to the scattering rate by providing an extra scattering channel but also increases the carrier concentration and hence reduces further the carrier scattering time. For bilayer graphene subject to a ≈3 Vnm$^{-1}$ bias field one finds that the gap energy is ≈0.28 eV [7]. A 1.55 µm (194 THz) NIR field creates carriers of 0.8 eV. Thus, one would expect a contribution from impact ionization. The interaction Hamiltonian for the process reads

$$\hat{H}_{\mathrm{ii}} = \frac{1}{2}\int d^2k\, d^2k'\, d^2q\, V_{\mathbf{q}} \left\{ \hat{e}^\dagger_{\mathbf{k+q}} \hat{e}^\dagger_{\mathbf{k'-q}} \hat{h}^\dagger_{\mathbf{k'}} \hat{e}_{\mathbf{k}} + \hat{e}^\dagger_{\mathbf{k}} \hat{e}_{\mathbf{k+q}} \hat{e}_{\mathbf{k'-q}} \hat{h}_{\mathbf{k'}} \right\}\Big|_{\mathbf{q}\neq 0}, \tag{8}$$

with $V_{\mathbf{q}}$ given in Supplementary Eq. (5) and the matrix elements reading [3, 8, 10]

$$|\langle \hat{H}_{\mathrm{ii}} \rangle|^2 = |V_{\mathbf{q}}|^2 [1 + \cos(\theta_{\mathbf{k+q}} - \theta_{\mathbf{k}})][1 - \cos(\theta_{\mathbf{k'-q}} - \theta_{\mathbf{k'}})]. \tag{9}$$

As before we will assume that the excited carrier density is low and, hence, the valence band is fully occupied. The scattering rate for a given occupied momentum state **k** reads

$$\Gamma^{\mathrm{ii}}_{\mathbf{k}} = \frac{2}{A_0} \frac{4\pi}{\hbar} \int d^2k'\, d^2q\, |\langle \hat{H}_{\mathrm{cc}} \rangle|^2 \delta\big[E(\mathbf{k+q}) + E(\mathbf{k'-q}) - E(\mathbf{k}) - E(\mathbf{k'}) + E_{\mathrm{g}}(\mathbf{k'})\big], \tag{10}$$

where the first prefactor is the density of valence band electrons ($A_0 = 5.24$ Å$^2$ is the area of the elementary unit cell which contains 2 electrons). The $\delta$-function enforces energy conservation with $E_{\mathrm{g}}(\mathbf{k'})$ denoting the energy required for an interband transition at momentum, $\mathbf{k'}$. Evaluating the above equation leads to an estimate for the impact ionization time of ≈15.50 ps, which leads to an increased carrier concentration per half-cycle of ≈3%. This time is somewhat longer than in conventional semiconductors. This is due to the difficulty of conserving both energy and momentum when the electrons have linear dispersion. As a result the phase space for scattering is smaller and hence the scattering time is longer. In fact, Supplementary Ref. [12] predicts impact ionization times as high as 100ps for certain temperatures and carrier concentrations. Furthermore, the presence of a gap in the BLG system we consider here will further reduce the phase space and hence one would expect an additional increase in the scattering time as compared to single-layer graphene [12, 13]. Finally, and as expected, the rate is lower than the carrier-carrier scattering rate, for the same reason that it is harder to conserve both energy and momentum in the case where energy, $E_g(\mathbf{k}')$, is lost to the band transition. The effect of impact ionization on the carrier-carrier scattering rate is shown in Supplementary Fig. 2.

**Carrier-phonon scattering rate:**

Only optical phonons close to the Γ and K points contribute significantly to the scattering of carriers with energy in the hundreds of meV [8]. Γ and K point phonons are responsible for intra- and inter- valley phonon scattering, respectively. Scattering by optical phonons can proceed via one of the two pathways; a carrier of momentum **k** absorbs a phonon of momentum **q**, scattering into the state **k** + **q** (+ve pathway), or a carrier of momentum **k** emits a phonon of momentum **q**, scattering into the state **k** − **q** (−ve pathway). At room temperature the optical phonon density is low and hence the −ve pathway is dominant. The interaction Hamiltonians for these two pathways read

$$\widehat{H}_{\text{cp}}^{+,(\mu,\nu)} = \frac{1}{2}\int d^2\text{k}\, d^2\text{q}\, g_\mathbf{q}^{+,(\mu,\nu)} \{\hat{e}_{\mathbf{k}+\mathbf{q}}^\dagger \hat{e}_\mathbf{k} \hat{p}_\mathbf{q} + \hat{h}_{\mathbf{k}+\mathbf{q}}^\dagger \hat{h}_\mathbf{k} \hat{p}_\mathbf{q}\}, \qquad (11\text{a})$$

$$\widehat{H}_{\text{cp}}^{-,(\mu,\nu)} = \frac{1}{2}\int d^2\text{k}\, d^2\text{q}\, g_\mathbf{q}^{-,(\mu,\nu)} \{\hat{e}_{\mathbf{k}-\mathbf{q}}^\dagger \hat{e}_\mathbf{k} \hat{p}_\mathbf{q}^\dagger + \hat{h}_{\mathbf{k}-\mathbf{q}}^\dagger \hat{h}_\mathbf{k} \hat{p}_\mathbf{q}^\dagger\}, \qquad (11\text{b})$$

where $g_\mathbf{q}^{\pm,(\mu,\nu)}$ are the carrier-phonon coupling constants with μ indicating the K or Γ point phonon and $\nu$ indicating the longitudinal (LO) or transverse (TO) phonons. Note that flexural phonons (ZO) are strongly suppressed for graphene supported on a substrate [14] and hence their contribution to the decay rate is negligible. The matrix elements read [8]

$$\left|\langle \hat{H}_{cp}^{\pm,(\Gamma,LO)}\rangle\right|^2 = \langle\left(g_\mathbf{q}^{\pm,(\Gamma,LO)}\right)^2\rangle\left[1 + \cos(\theta_\mathbf{k} + \theta_{\mathbf{k}\pm\mathbf{q}} - 2\theta_\mathbf{q})\right], \tag{12a}$$

$$\left|\langle \hat{H}_{cp}^{\pm,(\Gamma,TO)}\rangle\right|^2 = \langle\left(g_\mathbf{q}^{\pm,(\Gamma,TO)}\right)^2\rangle\left[1 - \cos(\theta_\mathbf{k} + \theta_{\mathbf{k}\pm\mathbf{q}} - 2\theta_\mathbf{q})\right], \tag{12b}$$

$$\left|\langle \hat{H}_{cp}^{\pm,(K,LO)}\rangle\right|^2 = \langle\left(g_\mathbf{q}^{\pm,(K,LO)}\right)^2\rangle\left[1 + \cos(\theta_\mathbf{k} + \theta_{\mathbf{k}\pm\mathbf{q}})\right], \tag{12c}$$

$$\left|\langle \hat{H}_{cp}^{\pm,(K,TO)}\rangle\right|^2 = \langle\left(g_\mathbf{q}^{\pm,(K,TO)}\right)^2\rangle\left[1 + \cos(\theta_\mathbf{k} + \theta_{\mathbf{k}\pm\mathbf{q}})\right], \tag{12d}$$

with the trigonometric factors a result of the chirality of the carriers. Once again, we will assume that the excited carrier density is low, the occupation of each momentum state is small and treat Pauli blocking as negligible. Furthermore, as the variation in phonon energy with momentum is small we will assume that the energy of the phonon and the electron-phonon coupling constants are constant for all momenta, **q** (Debye approximation). Hence, the scattering rate for a given occupied momentum state **k,** reads

$$\Gamma_\mathbf{k}^{cp} = \frac{2\pi A_0}{\hbar}\sum_{\mu,\nu}\int d^2q \left|\langle \hat{H}_{cp}^{+,(\mu,\nu)}\rangle\right|^2 \delta[E(\mathbf{k}+\mathbf{q}) - E(\mathbf{k}) - \hbar\omega^{(\mu,\nu)}]n^{(\mu,\nu)}$$

$$+ \frac{2\pi A_0}{\hbar}\sum_{\mu,\nu}\int d^2q \left|\langle \hat{H}_{cp}^{-,(\mu,\nu)}\rangle\right|^2 \delta[E(\mathbf{k}-\mathbf{q}) - E(\mathbf{k}) + \hbar\omega^{(\mu,\nu)}][n^{(\mu,\nu)} + 1], \tag{13}$$

where $n^{(\mu,\nu)}$ is the Bose-Einstein distribution for the $\nu$ polarized μ point phonon and the δ-function enforces energy conservation. $A_0 = 5.24$ Å$^2$ is the area of the elementary unit cell and the phonon energies [15] are $\hbar\omega^{(\Gamma,LO)} = 196$ meV, $\hbar\omega^{(\Gamma,TO)} = 194$ meV, for the Γ point phonons and $\hbar\omega^{(K,LO)} = 148$ meV, $\hbar\omega^{(K,TO)} = 157$ meV, for the K point phonons and the coupling constants [16] read $\langle(g^{\pm,(\Gamma,\nu)})^2\rangle = 0.0405$ eV$^2$, $\langle(g^{\pm,(K,\nu)})^2\rangle = 0.0994$ eV$^2$. We take the temperature of the phonon distribution to be 300 K (although this will not affect the final result since the optical phonon population at room temperature is far below one). The carrier-phonon rate is independent of carrier concentration and hence independent of the NIR field intensity. The above values lead to a carrier lifetime of ≈0.65ps.

**Total scattering rate:**

The total scattering rate is given by $\Gamma_\mathbf{k}^T = \Gamma_\mathbf{k}^{cc} + \Gamma_\mathbf{k}^{ii} + \Gamma_\mathbf{k}^{cp}$. Supplementary Fig. 3 shows the total relaxation time as a function of the NIR field strength and includes the contribution from the extra carriers created by impact ionization. The carrier lifetime exhibits an inverse proportionality to the NIR field strength for high NIR field strengths. In this regime carrier-carrier scattering is dominant. As the NIR field strength is reduced the lifetime flattens off asymptotically to the values of the carrier-phonon scattering time. In this regime the carrier-carrier scattering rate is negligible compared to the carrier-phonon scattering rate and hence gives no contribution to the relaxation rate. For NIR field intensities of $3.2 \times 10^{-4}$ mW/μm$^2$ (comparable to current HSG experiments) the total carrier lifetime is found to be ≈0.57 ps.

**SUPPLEMENTARY NOTE 4: Derivation of the cutoff law**

The sideband spectrum can be found from the Fourier transform of the polarization, which is given by the expectation value of the dipole moment operator [Eq. (4) in the main text], which in turn, can be computed from the electron-hole pair wave function whose time evolution is given by Eq. (2) in the main text. The formal solution to Eq. (2) leads to a sideband spectrum

$$\mathbf{P}(\omega) = -\frac{2i}{\pi\hbar}\int d^3k \int dt \int_0^\infty d\tau\, \mathbf{d}_{vc}^*[\widetilde{\mathbf{k}}(t)] e^{iS(\widetilde{\mathbf{k}},t,\tau)} \mathbf{d}_{vc}^*[\widetilde{\mathbf{k}}(t)] \cdot \mathbf{F}_{ex}(t), \qquad (14)$$

with the quasi-classical action

$$S(\widetilde{\mathbf{k}},t,\tau) = -\frac{1}{\hbar}\int_{t-\tau}^{t} dt''\, \Delta E[\widetilde{\mathbf{k}}(t)] + \omega t, \qquad (15)$$

where $\Delta E[\widetilde{\mathbf{k}}(t)]$ is given in Eq. (3) in the main text. Note that an extra factor of four appears in Supplementary Eq. (14) as a result of the spin and valley degeneracy of the bands as one obtains equal contributions from each. For parabolic band semiconductors the expression in Supplementary Eq. (14) leads to Gaussian-like integrals that can be evaluated via a saddle-point or stationary phase approximation [17-19]. Since the BLG bands are not of quadratic form they do not lend themselves to the traditional evaluation techniques and hence cannot be evaluated directly. In order to obtain an analytically tractable expression, a number of simplifications need to be made.

For an excitation laser of 1.55 μm, a wavelength located at the centre of the telecomm C-band, the dominant contribution to the sideband spectrum comes from transitions between the highest valence band, $E_v[\widetilde{\mathbf{k}}(t)]$, and the lowest conduction band, $E_c[\widetilde{\mathbf{k}}(t)]$. In this region of the Brillouin zone the bands are approximately linear so one can expand the bands to first order about the excitation point, $\mathbf{k}_{ex}$,

$$E_c[\widetilde{\mathbf{k}}(t)] - E_v[\widetilde{\mathbf{k}}(t)] \approx E_g + 2\hbar\widetilde{\mathbf{v}}_{f,vc} \cdot (\widetilde{\mathbf{k}}(t) - \mathbf{k}_{ex}), \qquad (16)$$

where

$$E_g = E_c[\mathbf{k}_{ex}] - E_v[\mathbf{k}_{ex}], \qquad (17)$$

and

$$\widetilde{\mathbf{v}}_{f,vc} = \frac{\hbar v_f^2 \mathbf{k}_{ex}(E_v[\mathbf{k}_{ex}] - E_c[\mathbf{k}_{ex}])}{2 E_v[\mathbf{k}_{ex}] E_c[\mathbf{k}_{ex}]} \left[1 - \frac{\gamma^2 + V^2}{\sqrt{v_f^2 \mathbf{k}_{ex}^2(\gamma^2 + V^2) + \gamma^4}}\right]. \qquad (18)$$

For the parameters given in the main text one finds that the effective Fermi velocity is $|\tilde{\mathbf{v}}_{f,vc}| = 5.90 \frac{eV\text{Å}}{\hbar}$ (close to the value of $6.46 \frac{eV\text{Å}}{\hbar}$ for monolayer graphene). Note that, owing to the rotational symmetry of the bands, the excitation resonance will consist of a circle of radius $|\mathbf{k}_{ex}|$ centred on the Dirac point. Essentially, the bands look like a shifted version of the mono-layer graphene bands with a modified Fermi velocity given by Supplementary Eq. (18).

By substituting the expression in Supplementary Eq. (16) into the quasi-classical action in Supplementary Eq. (15) and then minimizing it with respect to each variable, one can find the classical equations of motion for the system. For a THz field polarized in the x-direction one finds

$$E_g + 2\widetilde{U}_{p,vc}\tilde{\mathbf{v}}_{f,vc} \cdot \left(\frac{\widetilde{\mathbf{k}}(t-\tau) - \mathbf{k}_{ex}}{\tilde{\mathbf{v}}_{f,vc} \cdot \boldsymbol{\alpha}}\right) = \hbar\omega_{ex} - i\Gamma, \tag{19a}$$

$$E_g + 2\widetilde{U}_{p,vc}\tilde{\mathbf{v}}_{f,vc} \cdot \left(\frac{\widetilde{\mathbf{k}}(t) - \mathbf{k}_{ex}}{\tilde{\mathbf{v}}_{f,vc} \cdot \boldsymbol{\alpha}}\right) - \hbar\omega_{ex} + i\Gamma = \hbar\omega_s, \tag{19b}$$

$$2\tilde{\mathbf{v}}_{f,vc}\tau = 0, \tag{19c}$$

where $\boldsymbol{\alpha} = \frac{e\mathbf{F}_{THz}}{\hbar\omega_{THz}}$ is the maximum momentum the electron-hole pair can obtain from the THz field, $\hbar\omega_s = \hbar\omega - \hbar\omega_{ex}$ is the sideband energy centred around the optical excitation frequency and $\widetilde{U}_{p,vc} = \frac{e\tilde{\mathbf{v}}_{f,vc}\cdot\mathbf{F}_{THz}}{\omega_{THz}}$ is the linear ponderomotive energy associated with particle motion in each of the two coordinate directions. The classical equations give the constraint conditions for a fully classical electron-hole pair to produce sidebands. Supplementary Eq. (19a) states that the electron-hole pairs are created at time $t - \tau$ with an energy equal to the excitation field energy. Supplementary Eq. (19b) states that the sideband energy is given by the difference in the energy of the electron-hole pair at time $t$ and $t - \tau$. Supplementary Eq. (19c) states that the electron-hole pair must be at the same location to recombine, which occurs only when the transit time, $\tau$, is zero for linear dispersion bands. This is an indication that the particles behave like massless Dirac particles because, for non-zero transit times, the classical separation grows linearly at twice the Fermi velocity. For classical particles the constraints imposed by Supplementary Eqs. (19a-c) cannot be satisfied for non-zero $\tau$ and hence one would not expect to see sidebands. However, for quantum particles there may be non-vanishing overlap of the electron and hole wavefunctions and hence the particles do not have to be in the same location to recombine. Following Ref. [20], we relax the last constraint in Supplementary Eq. (19c) and maximise the sideband energy under the conditions of the remaining two. This leads to

$$\hbar\omega_s = 2\widetilde{U}_{p,vc}(\sin[\omega_{THz}(t-\tau)] - \sin[\omega_{THz}t]), \tag{20}$$

which is maximized for $t = \frac{3\pi}{2}, \tau = \pi$ and minimized for $t = \frac{\pi}{2}, \tau = \pi$. Thus, one expects a symmetric plateau about the excitation frequency with cutoffs at $\pm 4\widetilde{U}_{p,vc}$. The maximum/minimum energy sideband is given by electron-hole pairs that are excited at the point of minimum/maximum THz field and accelerated/decelerated for a half cycle. For electron-hole pairs that travel for longer, the THz field will reverse direction part way through the electron-hole pair's motion decelerating/accelerating the particles resulting in an energy change that is smaller than the maximum. $4\widetilde{U}_{p,vc}$ represents the maximum energy the electron-hole pair can gain or lose via interactions with the THz field.

**SUPPLEMENTARY NOTE 5: The Berry connection**

One important difference between BLG and traditional semiconductors is the appearance of the Berry connection, $\mathbf{A}_i[\widetilde{\mathbf{k}}(t)] = e\langle i, \widetilde{\mathbf{k}}(t)|\boldsymbol{\nabla}_\mathbf{k}|i, \widetilde{\mathbf{k}}(t)\rangle$, where the index, $i$, refers to one of the four energy bands. This term is a result of the non-trivial geometry of the bands and, as the dipole moments are dependent on **k**, is required to maintain the gauge invariance of the optical response. For circularly a polarized THz field, where trajectories of the driven electrons perform elliptic orbits around the Dirac point, or for electrons in a magnetic field, which undergo cyclotron orbits, the geometric phase leads to a number of non-trivial effects [21-24]. However, in the present case, with a linearly polarized THz field in the absence of a magnetic field, the contribution from the geometric phase vanishes.